%% file: main.tex
\documentclass[prb,twocolumn,showpacs,preprintnumbers,amsmath,amssymb,floatfix]{revtex4}

\usepackage{graphicx}
\usepackage{dcolumn}
\usepackage{bm}
\usepackage{graphicx,rotating}
\usepackage{epsfig}

\begin{document}

\title{Origins of low- and high-pressure discontinuities of $T_{c}$ in niobium}

\author{Ma\l{}gorzata Wierzbowska$^{1}$\footnote[3]{To whom correspondence should
be addressed (wierzbom@ts.infn.it)}, Stefano de Gironcoli$^{1,2}$ and Paolo Giannozzi$^{1,3}$} 

\affiliation{$^{1}$CNR-INFM DEMOCRITOS National Simulation Center, 
via Beirut 2--4, 34014 Trieste, Italy, \\
$^{2}$International School for Advanced Studies (S.I.S.S.A),
Via Beirut 2-4, 34014 Trieste, Italy \\
$^{3}$Scuola Normale Superiore, Piazza dei Cavalieri 7, I-56126 Pisa, Italy  
}%

\date{\today}

\begin{abstract}
The discontinuities of $T_{c}$ in Niobium under pressure 
are examined by means of the pseudopotential plane-wave implementation 
of the electron-phonon coupling  calculated from density-functional perturbation theory. 
Both low- and high-pressure discontinuities of $T_{c}$ 
have their origin in the Kohn anomalies and are caused by the low-frequency phonons, but the mechanism leading to the discontinuities is different in the two cases.
The low-pressure anomaly  is associated with a global decrease of the nesting factor in the whole Brillouin Zone and not to a visible change in the band 
structure.
The high-pressure anomaly is instead connected with a well-pronounced change in the band structure.
\end{abstract}

\maketitle

\input{introduction}

\input{def}

\input{matrix}

\input{bulk}

\input{Fermi}

\input{phonons}

\input{discussion}

\input{summary}

\begin{acknowledgments}
We would like to thank Andrea Dal Corso for useful discussion. 
Research in Trieste was supported through by INFM's
 ``Iniziativa Trasversale Calcolo Parallelo''.

\end{acknowledgments}

\begin{appendix}
\input{test}

\end{appendix}


\input{refs}
\end{document}

%% file: introduction.tex
\section{Introduction}

Niobium is a superconductor with a quite high critical temperature, 
$T_{c}$=~9.25~K, for a simple metal. The experiments under pressure 
by Struzhkin {\em et al.} \cite{Struzhkin} show discontinuities of $T_{c}$ 
at about 5~GPa and at 50-60~GPa.
The low-pressure discontinuity manifests itself as an increase of $T_{c}$ 
by about 1~K. The high-pressure anomaly is associated with a decrease 
of the critical temperature. To date, the nature of these pressure-induced 
discontinuities is not clear. Previous theoretical 
studies\cite{Ostanin,Ma,Anderson-1,Anderson-2} agree in attributing the
high-pressure anomalies to some visible change in the band structure.
The low-pressure discontinuity of $T_{c}$, however, remains mysterious.
The goal of this work is to give more information about the nature and 
the origin of the anomalous behavior of $T_{c}$ in niobium under pressure. 
In particular, we want to understand the role of Kohn
anomalies\cite{anomalies} of the phonon spectra. Kohn anomalies are known 
to drastically change the critical temperature in superconductors and are 
believed to play a role in all body-centered cubic (bcc) metals 
Nb, Mo, V, Ta.\cite{anomalies-exp}

Therefore, we look to the details of the electronic structure and dynamical 
properties of Nb at eight pressures in the range from -16 GPa to 78 GPa. 
We show that  Kohn anomalies are responsible for both discontinuities,
but the origin of the low-pressure Kohn anomaly is different 
from that of the high-pressure one.
Both can be identified  by a closer study of the Fermi-surface
nesting and of the band structure. 

The accurate calculation of the electron-phonon coupling $\lambda$ and of the
spectral function $\alpha^{2}F$ is crucial for our problem. To this end
we use density-functional perturbation theory\cite{DFPT,Review} (DFPT)
in a pseudopotential plane-wave approach. The Eliashberg function and the nesting
factor require an integration of the double delta over the Fermi surface, which
needs to be done with a high numerical accuracy.
We give a few advices for an efficient calculation of 
the electron-phonon coupling. Some of the technical details, however,
can be used in general for calculations of other properties which 
require an accurate numerical integration with the delta function.

This paper is organized as follows: In the next Section we remind the 
physical definitions and give some details of the calculation of 
electron-phonon interaction coefficients using Vanderbilt's ultrasoft 
pseudopotentials.\cite{Vanderbilt} 
In Sec. III, we give the technical details (Subsec. A) and present results 
for several properties under pressure: 
the lattice constant and bulk modulus (Subsec. B), 
the band structure and Fermi surface (Subsec. C), 
the phonon frequencies and linewidths (Subsec. D), and
the Eliashberg function and electron-phonon coupling constant (Subsec. E).
In Sec. IV, we discuss the origin of the anomalies, and we summarize in Sec. V.
In the Appendix, we give numerical details for the 
calculation of the Eliashberg function.

%% file: def.tex
\section{Electron-phonon coupling}

\subsection{Definitions}

The Hamiltonian for the electron-phonon interaction is given 
in second quantization by
\begin{equation}
H_{el-ph} = \sum_{{\bf k}{\bf q }\nu} 
g_{{\bf k}+{\bf q},{\bf k}}^{{\bf q}\nu ,mn} \; 
c_{{\bf k}+{\bf q}}^{\dagger m}c_{{\bf k}}^{n} \; (b_{-{\bf q}\nu}^{\dagger} + b_{{\bf q}\nu})
\end{equation}

\noindent
where $c_{{\bf k}+{\bf q}}^{\dagger m}$ and $c_{{\bf k}}^{n}$ are the creation and 
the annihilation operators for the quasiparticles with  energies
$\varepsilon_{{\bf k}+{\bf q},m}$ and  $\varepsilon_{{\bf k},n}$ 
in bands $m$ and $n$ with wavevectors ${\bf k}+{\bf q}$ and ${\bf k}$, 
respectively;
$b_{{\bf q}\nu}^{\dagger}$ and  $b_{{\bf q}\nu}$ are the creation 
and the annihilation operators for
phonons with energy $\omega_{{\bf q}\nu}$ and wavevector ${\bf q}$;
the matrix element $g_{{\bf k}+{\bf q},{\bf k}}^{{\bf q}\nu, mn}$  
describes the electron-phonon coupling. 
The coupling constants $g_{{\bf k}+{\bf q},{\bf k}}^{{\bf q}\nu, mn}$ 
define the spectral function, $\alpha^{2}F( \omega )$, its first reciprocal
momentum, $\lambda$, and the superconducting electron-phonon coupling constant,
$\lambda_{{\bf q}\nu}$, by the following set of equations:
\begin{eqnarray}
\alpha^{2}F( \omega) &=& \frac{1}{N(\varepsilon_{F})} \sum_{mn}\sum_{{\bf q}\nu}
\delta (\omega - \omega_{{\bf q}\nu})\sum_{\bf k} 
|g_{{\bf k}+{\bf q},{\bf k}}^{{\bf q}\nu, mn}|^{2} \nonumber \\ 
&& \times \delta (\varepsilon_{{\bf k}+{\bf q},m}-\varepsilon_{F})
\delta (\varepsilon_{{\bf k},n}-\varepsilon_{F}), \label{a2F} \\
\lambda & = & 
2 \int \frac{\alpha^{2}F( \omega)}{\omega} d\omega =  \sum_{{\bf q}\nu} \lambda_{{\bf q}\nu}, \label{lambda} \\
\lambda_{{\bf q}\nu} & = & \frac{2}{N(\varepsilon_{F})\omega_{{\bf q}\nu}} \sum_{mn} \sum_{\bf k} 
|g_{{\bf k}+{\bf q},{\bf k}}^{{\bf q}\nu,mn} |^{2} \nonumber \\ 
 &&  \times \delta ( \varepsilon_{{\bf k}+{\bf q},m} - \varepsilon_{F})
\delta ( \varepsilon_{{\bf k},n} - \varepsilon_{F} ). \label{dirlam}
\end{eqnarray}

\noindent
The quantity $N(\varepsilon_{F})$ is the density of states at the Fermi energy, 
$\varepsilon_{F}$, per both spins.

We introduce, after Allen,\cite{AllenGamma} the phonon linewidth 
$\gamma_{{\bf q}\nu}$:
\begin{eqnarray}
\gamma_{{\bf q}\nu} & = & 2 \pi \omega_{{\bf q}\nu} \sum_{mn}\sum_{{\bf k}} 
|g_{{\bf k}+{\bf q},{\bf k}}^{{\bf q}\nu, mn}|^{2} \; \nonumber \\
&& \times \delta (\varepsilon_{{\bf k}+{\bf q},m}-\varepsilon_{F}) 
\delta (\varepsilon_{{\bf k},n}-\varepsilon_{F}) \label{eq5} 
\end{eqnarray}

\noindent
which enters the Eliashberg function, $\alpha^{2}F$, and the
electron-phonon coupling constant, $\lambda_{{\bf q}\nu}$, as follows:

\begin{eqnarray}
\alpha^{2}F( \omega) & = & \frac{1}{2 \pi \; N(\varepsilon_{F})} \sum_{{\bf q}\nu}
\frac{\gamma_{{\bf q}\nu}}{\omega_{{\bf q}\nu}} \delta (\omega - \omega_{{\bf q}\nu}),
\label{a2Feq}   \\
\lambda_{{\bf q}\nu} & = & \frac{\gamma_{{\bf q}\nu}}
{\pi \; N(\varepsilon_{F}) \; \omega_{{\bf q}\nu}^{2}}. \label{lambdaeq}
\end{eqnarray}


\subsection{Matrix elements of the electron-phonon interactions}

Within DFPT\cite{DFPT,Review} the electron-phonon matrix elements 
can be obtained from the first-order derivative of the self-consistent 
Kohn-Sham\cite{DFT} (KS) potential, $V_{KS}$, 
with respect to atomic displacements, $\vec{u}_{s{\bf R}}$ for the 
$s-$th atom in lattice position ${\bf R}$, as:
\begin{equation}
g_{{\bf k}+{\bf q},{\bf k}}^{{\bf q}\nu, mn} = \left( \frac{\hbar}{2\omega_{{\bf q}\nu}} \right)^{1/2} 
\langle \psi_{{\bf k}+{\bf q},m} | \Delta V_{KS}^{{\bf q}\nu}  | \psi_{{\bf k},n} \rangle ,
\label{gg}
\end{equation}
where $\psi_{{\bf k},n}$ is the $n-$th valence KS orbital of wavevector
${\bf k}$ and
\begin{equation}
\Delta V_{KS}^{{\bf q}\nu} = \sum_{\bf R} \sum_{s} \frac{\partial V_{KS}}{\partial \vec{u}_{s{\bf R}}}
\cdot \vec{u}_{s}^{{\bf q}\nu} \frac{e^{i{\bf qR}}}{\sqrt{N}}
\label{dV}
\end{equation}
is the self-consistent first order variation of the KS potential, 
$N$ is the number of cells in the crystal, and
$\vec{u}_{s}^{{\bf q}\nu}$ is the displacement pattern for phonon mode 
$\vec{v}_{s}^{{\bf q}\nu}$:
\begin{equation}
\vec{u}_{s}^{{\bf q}\nu} = \frac{\vec{v}_s^{{\bf q}\nu}}{\sqrt{M_{s}}}
\end{equation}
The latter is obtained from the diagonalization 
of the dynamical matrix, $\varPhi^{\alpha\beta}_{ss'}({\bf q})$:
\begin{equation}
\sum_{s'\beta} \frac{\varPhi^{\alpha\beta}_{ss'}({\bf q})}{\sqrt{M_{s}M_{s'}}} 
v_{s'\beta}^{{\bf q}\nu}
 =  \omega_{{\bf q}\nu}^{2} v_{s\alpha}^{{\bf q}\nu}. \label{eq11}
\end{equation} 
$M_{s}$ is the mass of atom $s$, $\alpha, \beta$ denote cartesian coordinates.

%% file: matrix.tex
\subsection{Matrix elements with ultrasoft pseudopotentials}

The use of  ultrasoft (US) pseudopotentials (PPs)\cite{Vanderbilt} allows 
in many cases a significant reduction of the needed plane-wave kinetic 
energy cutoff, as compared with standard norm-conserving pseudopotentials.
\cite{HBS1,KB,MT} This enables a more efficient calculation, at the price of 
introducing additional terms originating from the augmentation charges 
employed in this scheme.\cite{Vanderbilt}
A detailed description of DFPT with US PPs has been given
elsewhere by Dal Corso.\cite{AndreaUS}  Here, we only briefly describe 
terms appearing in the electron-phonon coupling.

With US PPs the KS orbitals, $\psi_{{\bf k},n}$, satisfy 
a generalized eigenvalue problem
\begin{equation}
\left( -\frac{\nabla^{2}}{2} + V_{KS} - \varepsilon_{{\bf k},n} 
S \right) \psi_{{\bf k},n}=0
\end{equation}
where the overlap matrix, S, is given by 
\begin{eqnarray}
S({\bf r}_{1},{\bf r}_{2}) &=& \delta ({\bf r}_{1}-{\bf r}_{2}) +  \sum_{snm} q_{nm} \nonumber \\
&& \times \beta_{n}({\bf r}_{1}-{\bf R}_{s})
\beta_{m}^{\ast}({\bf r}_{2}-{\bf R}_{s}).
\label{SS}
\end{eqnarray}
The charge correction $q_{nm}$, in the above formula, is defined with the 
augmentation functions, 
$Q_{nm}({\bf r}-{\bf R}_{s})$, as follows
\begin{equation}
q_{nm} = \int d^{3}r \; Q_{nm}({\bf r}-{\bf R}_{s}), 
\label{charge}
\end{equation}
and the projector functions $\beta_{n}({\bf r}-{\bf R}_{s})$ are specific for the
type of atom at the position ${\bf R}_{s}$, and are obtained from the atomic calculations. 

The valence charge density is then computed as
\begin{eqnarray}
\rho({\bf r}) & = & \sum_{{\bf k},i} (|\psi_{i}({\bf r})|^{2} + \sum_{smn} Q_{mn}({\bf r}-{\bf R}_{s})
\langle \psi_{{\bf k},i} | \beta_{m} \rangle \langle \beta_{n}| \psi_{{\bf k},i} \rangle ) \nonumber \\
& = & \sum_{{\bf k},i} \int \int d^{3}r_{1}d^{3}r_{2} 
\psi_{{\bf k},i}^{\ast} ({\bf r}_{1}) K({\bf r}; {\bf r}_{1},{\bf r}_{2}) 
\psi_{{\bf k},i} ({\bf r}_{2}), 
\end{eqnarray}
where the sum over ${\bf k}$ and $i$ runs on occupied KS orbitals, 
the kernel
\begin{eqnarray}
K({\bf r}; {\bf r}_{1},{\bf r}_{2}) & = & \delta ({\bf r}-{\bf r}_{1})
\delta ({\bf r}-{\bf r}_{2}) + \sum_{snm} Q_{nm}({\bf r}-{\bf R}_{s}) \nonumber \\
&& \times  \beta_{n}({\bf r}_{1}-{\bf R}_{s})
\beta_{m}^{\ast}({\bf r}_{2}-{\bf R}_{s}) 
\end{eqnarray}
has been introduced for later convenience.
 
The KS selfconsistent potential in the US-PP scheme reads
\begin{eqnarray}
V_{KS} ({\bf r}_{1},{\bf r}_{2}) & = & V_{NL}({\bf r}_{1},{\bf r}_{2}) \nonumber \\
 && + \int d^{3}r  \; V_{eff}({\bf r})\; K({\bf r}; {\bf r}_{1},{\bf r}_{2}) .  
\label{VNL}
\end{eqnarray}
The effective potential, $V_{eff}$, contains the local,
the Hartree and the exchange-correlation (xc) terms 
\begin{equation}
V_{eff} ({\bf r}) = V_{loc} ({\bf r}) + \int d^{3}r_{1} \;
\frac{\rho({\bf r}_{1})}{|{\bf r}_{1}-{\bf r}|} + V_{xc} ({\bf r}), 
\end{equation}
while the nonlocal term generalizes the usual Kleinman-Bylander\cite{KB} 
form allowing several projectors for a given angular momentum component 
\begin{equation}
V_{NL} ({\bf r}_{1},{\bf r}_{2}) = \sum_{snm} D_{nm}^{0} \;
\beta_{n}({\bf r}_{1}-{\bf R}_{s}) \beta_{m}^{\ast}({\bf r}_{2}-{\bf R}_{s}).  
\end{equation}
When augmentation charges vanish ($Q_{nm}=0$) the above formulas
reduce to the standard norm-conserving formulation.

In order to generalize Eq.~(\ref{gg}) to the case of US PPs, one needs to 
compute 
first order perturbation theory in presence of overlap matrix, $S$, as below
\begin{equation}
g_{{\bf k}+{\bf q},{\bf k}}^{{\bf q}\nu, mn} = 
\langle \psi_{{\bf k}+{\bf q},m} | \Delta V_{KS}^{{\bf q}\nu}
- \varepsilon_{{\bf k},n} \Delta S | \psi_{{\bf k},n} \rangle , 
\end{equation}
where 
\begin{equation}
\Delta S = \sum_{\bf R} \sum_{s} \frac{\partial S}{\partial \vec{u}_{s{\bf R}}}
\cdot \vec{u}_{s}^{{\bf q}\nu} \frac{e^{i{\bf qR}}}{\sqrt{N}},
\end{equation}
and $\Delta V_{KS}^{{\bf q}\nu}$ is given by Eq.~(\ref{dV}).

The derivative of the Kohn-Sham potential is given in Ref.~[\onlinecite{AndreaUS}]
and for gradient-corrected functionals in Ref.~[\onlinecite{GGA}].





%% file: bulk.tex
\section{Results for niobium under pressure}

\subsection{Technical details}

The calculations of the ground-state electronic and vibrational
properties of Nb were performed using the Local-Density Approximation
(LDA) and an ultrasoft pseudopotential.
A kinetic energy cut-off of 45~Ry (270~Ry) was chosen for the expansion
into plane waves of the wavefunctions (density). Such high cut-offs were 
necessary to obtain accurate values for some "strategic" low-frequency
phonons, located mostly near the $\Gamma$-point. In fact, even small 
errors in this region of the spectrum lead to large relative errors in
the estimate of the  $\alpha^{2}F$ function  and of $\lambda$.

The integration over the Brillouin zone (BZ) requires special techniques
to account for the Fermi surface. We used the broadening technique proposed 
in Ref.~[\onlinecite{Paxton}] with a smearing parameter of 0.03 Ry 
(which was tested\cite{SdG-broad} to reproduce well
the experimental spectra).
The grids for the electronic BZ integration ({\bf k}-grid) 
and for the phononic BZ integration ({\bf q}-grid)
have been chosen according to the Monkhorst-Pack scheme.\cite{mesh}

Details of the numerical quadrature used to evaluate the double-delta
term appearing in Eq.(2), together with convergence tests,
are given in the Appendix.

All calculations were performed using the
{\sc quantum-espresso}\cite{espresso}  suite of codes.

\subsection{Structural properties}

Niobium in the body-centered cubic structure was studied at eight values 
of the lattice parameter from 6.34~a.u. to
5.64~a.u in steps of 0.1 a.u.  These lattice parameters correspond to
pressures ranging from -16.6~GPa to 78.4~GPa. The results are reported
in TABLE~\ref{units}. The calculated static equilibrium lattice constant
(zero-point motion and thermal effects not included) is about 6.14~a.u., 
slightly underestimating (as it usually happens within LDA)
the experimental value\cite{Ashcroft} of 6.24~a.u. The calculated bulk 
modulus at the theoretical equilibrium lattice is 192~GPa, versus a 
room-temperature 
experimental value\cite{bulkmod} of 170~GPa and a calculated value of
162~GPa at the experimental lattice parameter of 6.24 a.u.  
The calculated bulk modulus is very sensitive to the volume: it varies
by more than a factor three in the considered range of pressures.

\begin{table}
\begin{tabular}{cccc}
\hline \hline
\\[0.01mm]
  $\;\;\;$ $a$ $\;\;\;$ & $\;\;\;$  V/V$_{0}$ $\;\;\;$ & 
$\;\;\;$ $P$ $\;\;\;$ & $\;\;\;$ $B$ $\;\;\;$ \\[0.2cm]
\hline \\[0.01mm]
  6.34 &  1.10 & -16.6 & 134  \\
  6.24 &  1.05 & -9.5 & 162  \\
  6.14 &  1.00 &  -0.6 & 192  \\
  6.04 &  0.95 &  10.0 & 220  \\
  5.94 &  0.91 & 22.9 & 249  \\
  5.84 &  0.86 &  38.8 & 308  \\
  5.74 &  0.82 & 56.7 & 354  \\
  5.64 &  0.78 &  78.4 & 424  \\[0.05cm]
\hline \hline
\end{tabular}
\caption{\label{units} The lattice constant $a$ (in a.u.), the corresponding 
volume ratio V/V$_{0}$,
the pressure $P$ (in GPa), and bulk modulus $B$ (in GPa) calculated for BCC
niobium crystal at several pressures.
The experimental lattice constant\cite{Ashcroft} is 6.24 a.u.
and the bulk modulus\cite{bulkmod} is 170 GPa.}
\end{table}

%% file: Fermi.tex
\subsection{Band structure and Fermi surface}

\begin{figure} \epsfxsize=8cm \centerline{
\includegraphics[scale=0.33,angle=-90]{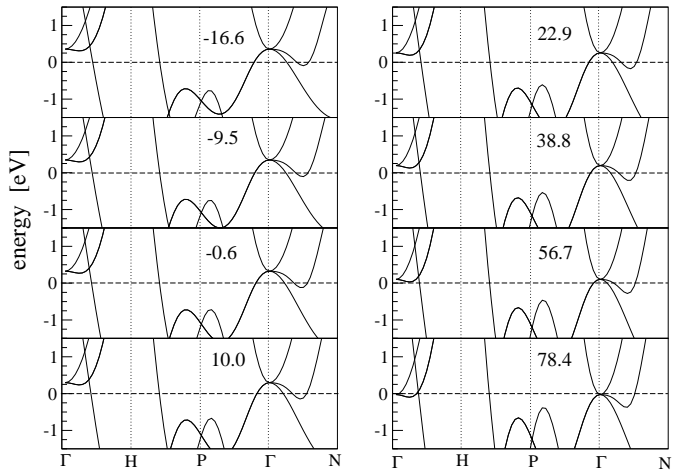}} \caption{\label{bands}
The band structures of niobium at several pressures (in GPa).}
\end{figure}

The evolution of the band structure as a function of pressure is presented in
FIG.~\ref{bands}, showing no qualitative change in the
electronic states at the Fermi surface up to about 38.8 GPa.
From 56.7~GPa to 78.4~GPa, we observe some changes along the $\Gamma$-H
and $\Gamma$-N lines. 

A 3D picture of the Fermi surface at ambient pressure and 56.7~GPa is drawn
in FIG.~\ref{Fermi}.  The lower energy band
forms the octahedron centered at the $\Gamma$-point. With increasing pressure,
this octahedron shrinks and it becomes surrounded
by six little ellipsoids when the previously described band-structure changes
on the $\Gamma$-H line appear.
Around N-point, the Fermi surface forms ellipsoids that are disconnected at
lower pressure and becomes connected by necks to the four neighboring
ellipsoids
above 56.7~GPa. In addition, a complicated open sheet structure, often
referred as "jungle gym", extends from $\Gamma$ to the H points in the BZ.

Our results are in good agreement with
the detailed studies of Anderson {\em et al.}~\cite{Anderson-1}
for the band structure and of Ref.~[\onlinecite{Anderson-2}] for the
Fermi surface.

\begin{figure}
\caption{\label{Fermi} Fermi surfaces from two bands (the lower energy
band in the left panels and the higher energy band in the right panels)
at ambient pressure (top panels) and at 56.7
GPa (bottom panels).  Pictures obtained with
the XCrySDen package.\cite{xcrysden}} \end{figure}

%% file: phonons.tex
\subsection{Phonon frequencies and linewidths}

\begin{figure*}
\epsfxsize=8cm
\centerline{
\includegraphics[scale=0.45,angle=-90]{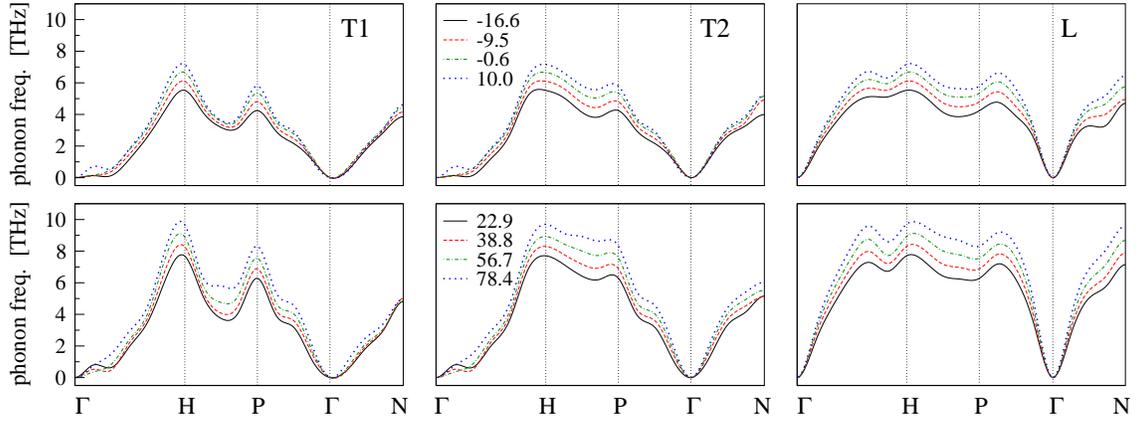}}
\caption{Phonon dispersions in niobium at several pressures (in GPa),
for the two transverse modes (left and medium panels, notation "T1" and "T2") 
and the longitudinal mode (right panels, notation "L");
the scale on the vertical axis is the same for all three modes.}
\label{ph-p}
\end{figure*}
\begin{figure*}
\epsfxsize=8cm
\centerline{
\includegraphics[scale=0.45,angle=-90]{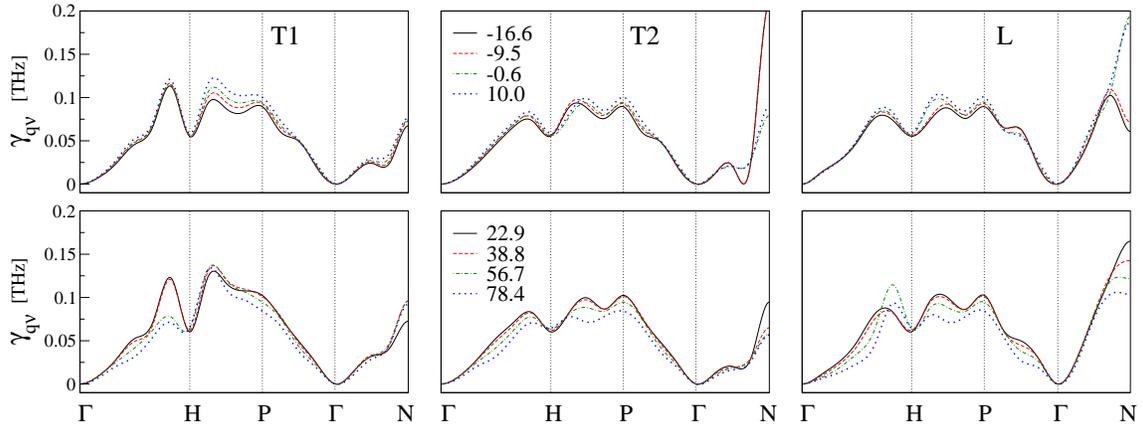}}
\caption{Phonon linewidths (in THz) in niobium 
at several pressures (in GPa).
The three modes of $\gamma_{{\bf q}\nu}$ 
correspond to the three phonon modes shown in FIG.~\ref{ph-p};
the scale on the vertical axis is the same for all three modes.}
\label{gam-p}
\end{figure*}
\begin{figure*}
\epsfxsize=8cm
\leftline{
\includegraphics[scale=0.45,angle=-90]{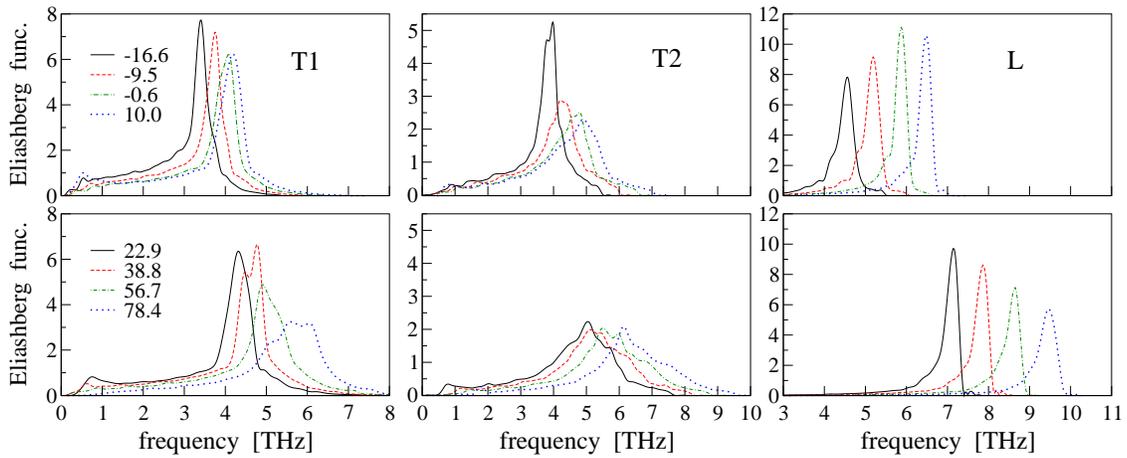}}
\caption{Eliashberg function, $\alpha^{2}F$, in niobium at several pressures (in GPa);
the scale on the horizontal axis is the same,
within a given mode, for all pressures.}
\label{a2F-p}
\end{figure*}

The phonon spectra are presented in FIG.~\ref{ph-p}. We observe
an overall increase in phonon frequencies with pressure,
especially at H, P and N high-symmetry points. 
Close to the $\Gamma$-point along the $\Gamma$-H symmetry
line, very low frequency phonon modes with an  anomalous
pressure dependence can be observed. At both the experimental and 
the calculated equilibrium lattice constants, the two
transverse modes (T1 and T2 in the following) display a
very flat dispersion close to  $\Gamma$-point. 
At variance with all other modes in the BZ, the T1 and T2 modes along $\Gamma$-H
line soften with pressure (between $\approx$10 to $\approx$55 GPa in
our calculations). 
This anomalous, non-monotonic, dispersion relation
eventually disappears and become a smooth curve
at the highest pressure we have considered.

The phonon linewidths $\gamma_{{\bf q}\nu}$ have a weak dependence upon 
pressure (see FIG.~\ref{gam-p}), with two important exceptions: 
i) at low pressure,
large variations in phonon linewidth occur for the T2 and L modes near the N
high-symmetry point, and ii) at high pressure (above $\approx$55 GPa),
a significant reduction in linewidth is observed in many parts of the BZ,
especially along the $\Gamma$-H direction.

\subsection{Eliashberg function and electron-phonon coupling constant}
\label{sec-lambda}

By integrating the calculated phonon linewidths and frequencies we can
obtain Eliashberg $\alpha^{2}F(\omega)$ function, Eq.\ (\ref{a2Feq}), that
we present in FIG.~\ref{a2F-p} for the three phonon branches separately.
As a general feature, all peaks in  $\alpha^{2}F(\omega)$ move to higher
frequency with pressure, as expected from the global positive frequency shift 
with pressure visible in FIG.~\ref{ph-p}.
For the T1 and T2 modes the height of the main peak decreases with pressure,
while for the L mode  the maximal height of the peak appears at the equilibrium
lattice constant.

Our calculated $\lambda$ and the density of states at the studied pressures 
are presented in TABLE~\ref{DOS-lambda}. 
The pressure behavior of the electron-phonon coupling constant 
is similar to what could be expected from the experimental pressure 
dependence for $T_{c}$:  $\lambda$ shows a positive jump at low pressure
and decreases significantly at high pressure.

\begin{table}
\begin{tabular}{ccccccc}
\hline \hline
\\[0.01mm]
$\;\;\;$  $P$ $\;\;\;$ & & $\;\;\;$ $N(\varepsilon_F)$ $\;\;\;$ 
&& $\;\;\;$ $\lambda$ $\;\;\;$ && 
$\;\;\;$ $\sim T_{c}^{exp}$  $\;\;\;$ \\[0.2cm]
\hline\\[0.01mm]
 -16.6 & &  11.41 && 1.91  && \-- \\
 -9.5  & &  10.80 &&  1.60 && \-- \\
 -0.6  & &  10.12 && 1.41  && 9.2 \\
 10.0   & &  9.69  &&  1.65 && 10.0 \\
 22.9  & &  9.16  &&  1.47 && 9.8   \\
 38.8  & &  8.55  &&  1.29 && 9.7  \\
 56.7  & &  7.71  &&  1.10 && 9.5 \\
 78.4  & &  6.55  &&  0.86 && 8.8 \\[0.05cm]
\hline \hline
\end{tabular}
\caption{The parameters of niobium under pressure, $P$ (in GPa):
the electronic density of states at the Fermi surface
$N(\varepsilon_F)$ (states per spin and per Ry),
the electron-phonon coupling constant $\lambda$,
and the experimental critical temperature $T_{c}^{exp}$
from Ref.~[\onlinecite{Struzhkin}].}
\label{DOS-lambda}
\end{table}

A good review of theoretical works on the electron-phonon coupling in Nb
at ambient pressure is given by Solanki {\it et al.} in Ref.~[\onlinecite{Solanki}],
where the reported values of the electron-phonon coupling constant range
from 0.59, obtained\cite{Anderson-1} from augmented plane-wave method
(APW), to 1.52, obtained\cite{lambda-1.52} later from the same  method.
More recently, Savrasov\cite{Sav-lambda} obtained a value of 1.26,
Bauer {\it et al.} \cite{Bauer} obtained a value of 1.33,
while our results for $\lambda$ at ambient pressure is 1.41.

We notice that the main reason for the spread in the
reported values of $\lambda$ is the difference in calculated values 
for $N(\varepsilon_F)$, entering the definition of $\lambda$ in the 
denominator (Eqs.~(\ref{a2F})-(\ref{dirlam})).
In fact a large value of 14.1 (states per spin and per Ry) for the DOS and a
small value of 0.59 for $\lambda$  have been
reported in Ref.~[\onlinecite{Anderson-1}], while
a value of 8.89 for the DOS and $\lambda$=1.52 have been reported in
Ref.~[\onlinecite{lambda-1.52}].
Consistent with this trend, our calculated DOS of 10.12 at ambient pressure
is somewhat lower than the DOS of 10.21 obtained by Savrasov,\cite{Sav-lambda},
while other authors obtained
$\sim 11$ (Ref.~[\onlinecite{Ostanin}]), 11.77 (Ref.~[\onlinecite{Singalas}]), 
and 9.84 (Ref.~[\onlinecite{Crabtree}]).

The experimental values of $\lambda$ obtained from the electronic
tunneling spectroscopy\cite{lambda-exp1,lambda-exp2} are 1.04 and 1.22,
while Haas-van Alphen data\cite{Crabtree} yield $\lambda=1.33$.


%% file: discussion.tex
\section{Discussion}


After examination of the phonons and the electron-phonon coupling,
we notice that the decrease of $\lambda$ at high pressure 
can be easily related to the  decrease of the $\alpha^{2}F(\omega)$ peak 
and to its shift toward higher frequencies for all modes.

The origin of the increase of $\lambda$ at low pressure between 
$\approx$0 GPa and $\approx$10 GPa is instead more difficult to trace.
We found out that it is  mainly determined by the anomalous dispersion 
of the T1 and T2 modes close to $\Gamma$-point (see FIG.~\ref{ph-p}) 
in the frequency region below 1~THz, 
that determines a low-frequency peak in the Eliashberg function of
the T1 mode above 10 GPa. This peak is instead absent at ambient pressure
and reappears for an expanded lattice only at about -16 GPa.

It has been shown \cite{Vonsovsky} for the Eliashberg model 
that the contribution to $T_{c}$ from acoustic modes close to the $\Gamma$-point
vanishes. For the low-frequency modes which are associated to Kohn anomalies,
however, we can expect important contribution to $T_{c}$ because phonon
softening may occur with a finite phonon linewidth.
Therefore, the region near $\Gamma$ can give a very large contribution 
to the electron-phonon coupling in niobium for all studied pressures 
(see TABLE~\ref{DOS-lambda}).


Let us now consider the band structure.
In order to explain low-pressure anomalies in $T_c$, 
Struzhkin {\em et al.}\cite{Struzhkin} proposed 
the existence of necks between the
ellipsoids around N and the "jungle gym" open sheet extending from
$\Gamma$ to H along the $\Gamma$-$\Sigma$ line at a pressure below 5~GPa,
and the disappearance of these features at the higher pressures.
Our calculations do not support this suggestion.
The detailed analysis of the Fermi surface reported by Ostanin
{\em et al.} \cite{Ostanin} gives results close to ours.
Previous theoretical
\cite{Solanki,Anderson-2,Neve,lambda-1.52,fermi-1,fermi-2} and
experimental \cite{Anderson-2,fermi-3,fermi-4} investigations of the
Fermi surface also did not detect any changes at low pressure.

One can, therefore, connect the high-pressure decrease 
in electron-phonon coupling to changes in the band structure, while
the origin of the low-pressure anomaly remains unclear.
Therefore, we need a different tool in order to detect tiny features
of the Fermi surface.

\begin{figure}
\epsfxsize=8cm
\includegraphics[scale=0.25,angle=-90.0]{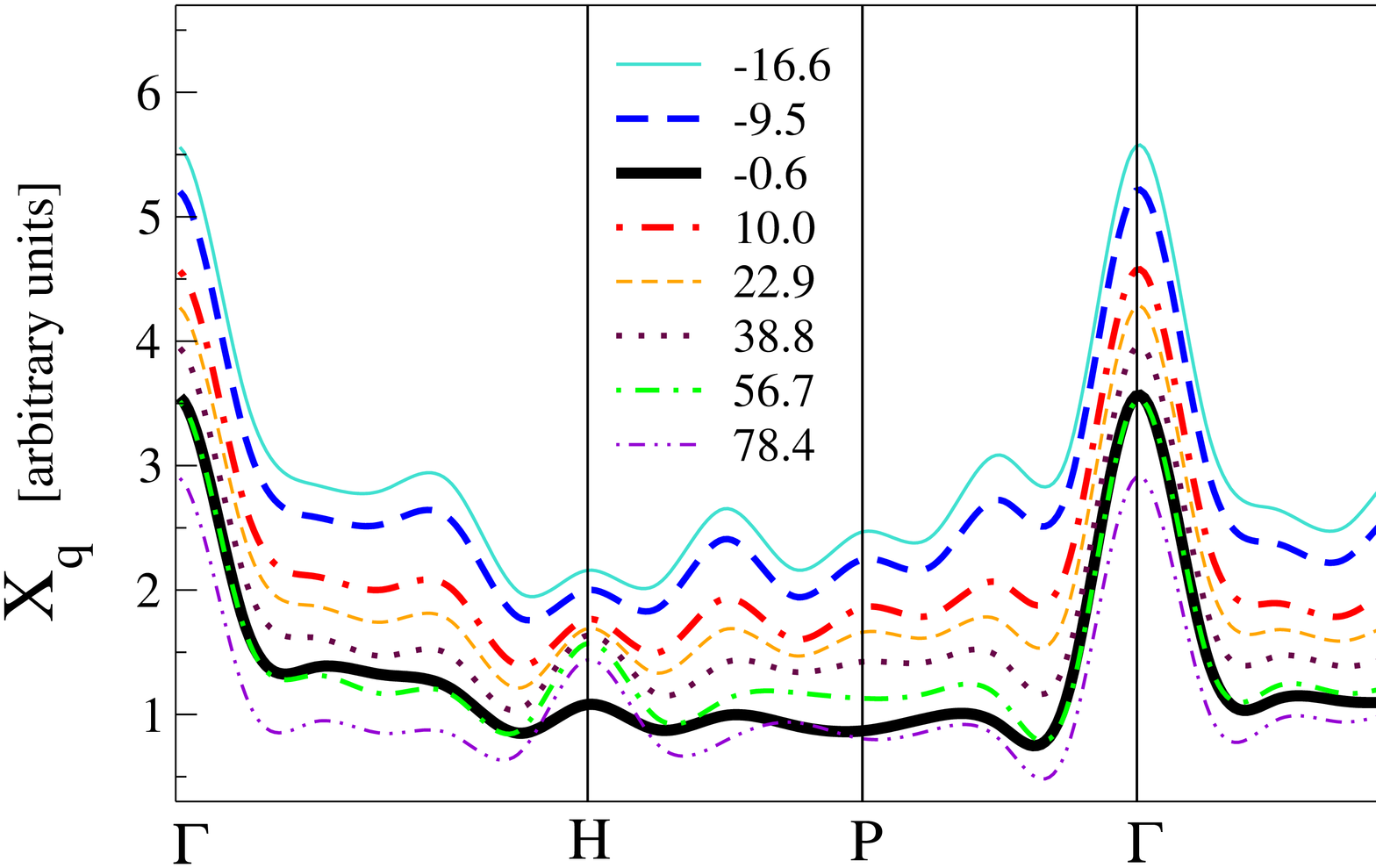}
\includegraphics[scale=0.25,angle=-90.0]{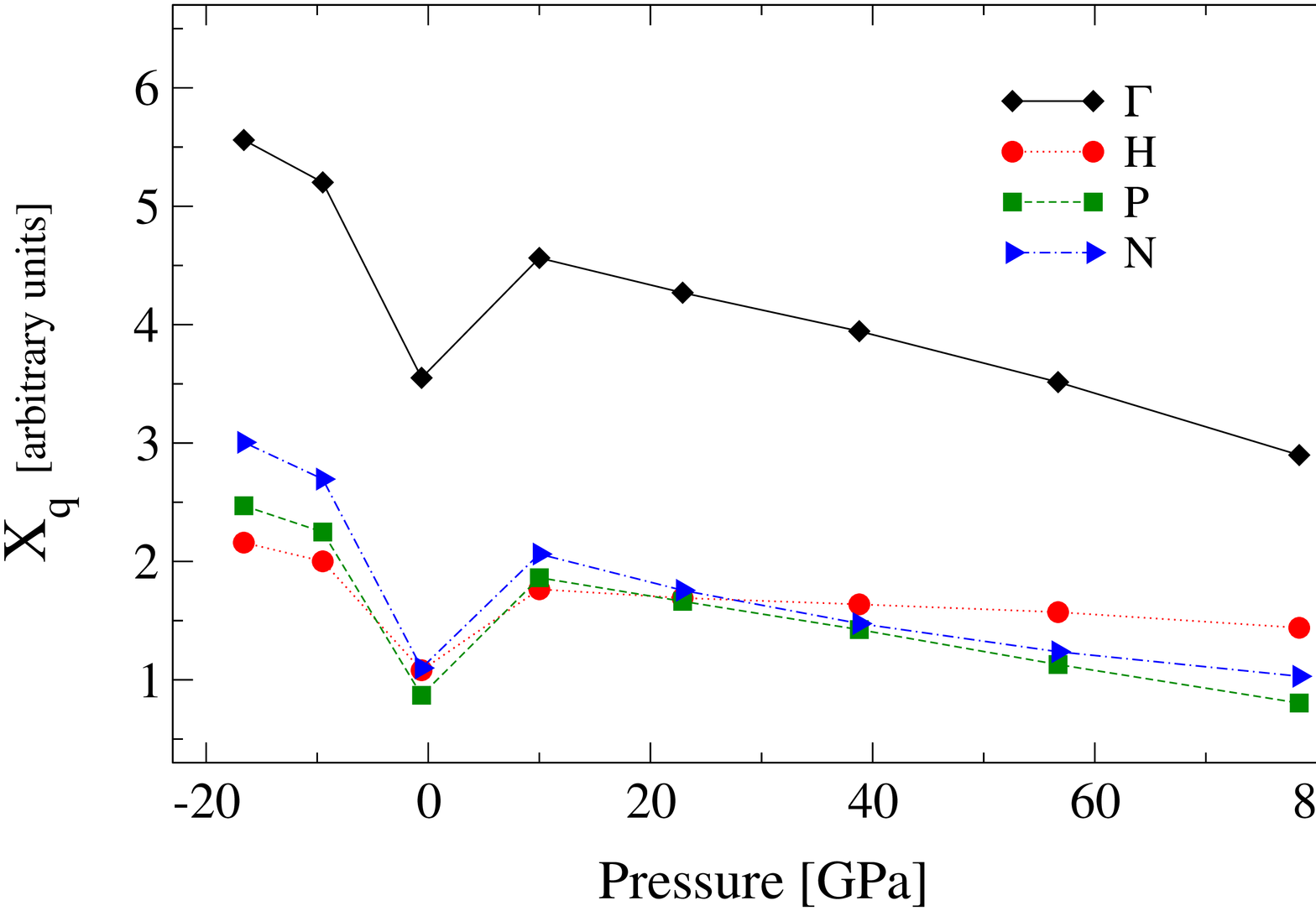}
\vspace{-0.4cm}
\caption{\label{nest} The nesting factor, $X_{\bf q}$,
of the Fermi surface in niobium at eight studied pressures (in GPa),
for selected high-symmetry lines (top panel) and 
high-symmetry points (bottom panel).}
\end{figure}
\begin{figure}
\caption{\label{nest-BZ} The isosurface of the nesting factor in the whole BZ,
$X_{\bf q}=1.0$ (arbitrary units), for niobium at pressures of -0.6 GPa (left panel)
and 10.0 GPa (right panel).
Pictures obtained with the XCrySDen package.\cite{xcrysden}} \end{figure}

We propose to look closer at Fermi surface nesting by plotting the
dispersion of the nesting factor:
\begin{equation}
 X_{\bf q}=\sum_{\bf k} \delta (\varepsilon_{{\bf k}}-\varepsilon_{F}) 
            \delta (\varepsilon_{{\bf k}+{\bf q}}-\varepsilon_{F}).
\end{equation} 
Large nesting factors correspond to large regions of the Fermi surface
being connected by the nesting vector, $\bf q$, and are expected to correspond
to large electron-phonon couplings.

In FIG.~\ref{nest}, the factor $X_{\bf q}$ is reported as a function of pressure.
This quantity has been computed numerically using the smearing
technique with the broadening of 0.03 Ry.  The maximal nesting takes place at
the $\Gamma$-point with much smaller maxima at the high-symmetry points
H, P and N, and in the middle of the lines: $\Gamma$-H, H-P, P-$\Gamma$
and $\Gamma$-N. The nesting factor decreases monotonically with
increasing pressure in the whole BZ except around ambient pressure
(0\--10~GPa). 
At the equilibrium lattice constant, a large damping
of the nesting factor moves the whole nesting curve  below its value
at 38.8~GPa and makes it even similar to the
curve drawn for the pressure of 56.7~GPa. 

The damping of the nesting factor close to ambient pressure 
occurs in the whole BZ, as we can see in FIG.~\ref{nest-BZ} in comparison
to nesting for a pressure of 10.0 GPa.   
This damping explains the jump in the total electron-phonon coupling constant 
which happens below and above the ambient pressure (see TABLE~\ref{DOS-lambda}).


%% file: summary.tex
\section{Summary}

We investigated the origin of the two discontinuities of the superconducting 
critical temperature, observed in Niobum  at low pressure, about 5~GPa, 
and at high pressure, about 60~GPa. \cite{Struzhkin}
For this purpose, we developed computational tools for the accurate
calculation of the electron-phonon coupling.

We find that the anomalous behavior of $T_{c}$ in Nb 
under pressure originates in Kohn anomalies close to the $\Gamma$-point 
in the BZ, so the measured discontinuities are caused by low-frequency phonons.
In agreement with previous authors,\cite{Ostanin,Ma,Anderson-1,Anderson-2} 
we find that the high-pressure discontinuity of $T_{c}$ is associated to
a visible change in the band structure. 
As for the low-pressure discontinuity, we notice that such anomaly 
shows up as a general decrease of the nesting  factor without any visible change 
in the shape of the Fermi surface. Such a decrease is uniform in the whole BZ, 
explaining why previous calculations, Refs. [\onlinecite{Ostanin,Ma}], did not detect any
anomaly in the electronic structure of Nb near ambient pressure.
The total electron-phonon coupling constant varies with pressure 
as expected from the measured critical temperatures. 

In conclusion, both discontinuities of $T_{c}$ in niobium, at low and high pressures, 
can be reproduced when the electron-phonon spectral function is calculated  accurately,
and the anomalies can be explained by the closer look into the details of 
the Fermi-surface nesting and the band structure.

%% file: test.tex
\section{Implementation details and test calculations}

The electron-phonon coupling constant $\lambda$ and the spectral
function $\alpha^{2}F(\omega)$ are defined by a double-delta
integration on the Fermi surface (Eqs.~(\ref{a2F}) and (\ref{dirlam})).
The accurate calculation of these integrands requires a very dense 
sampling in both the electronic ({\bf k}) and the phononic
({\bf q}) grids. One can use either the broadening technique
\cite{broadening} or the tetrahedron method.\cite{tetra} We choose
the former to perform the quadrature on the Fermi surface, and 
the latter to evaluate the electron-phonon and phonon densities of
states as functions of the vibrational frequencies.  In the broadening
scheme, a finite energy width is attributed to each state. For any 
function $f$ which has to be integrated with the double-delta, 
one can use the formula
\begin{eqnarray}
I  &=  &\int d{\bf k} \int d{\bf q} \; f({\bf k},{\bf q}) \; 
\delta(\epsilon_{\bf k}-\epsilon_{F}) \delta(\epsilon_{{\bf k}+{\bf q}}-\epsilon_{F})  
 \nonumber \\
&  \simeq & {\Omega_{BZ}^2\over N_k N_q} 
   \sum_{\bf k} \sum_{\bf q} f({\bf k},{\bf q}) \;
  \frac{1}{\sqrt{2\pi}\sigma} 
  exp \left( - \frac{(\epsilon_{\bf k}-\epsilon_{F})^{2}}{\sigma^{2}} \right) \nonumber \\
&&  \times \frac{1}{\sqrt{2\pi}\sigma}
  exp \left( - \frac{(\epsilon_{{\bf k}+{\bf q}}-\epsilon_{F})^{2}}{\sigma^{2}} \right),
\label{integral} 
\end{eqnarray} 
where $\sigma$ is the broadening, $N_k$ and $N_q$ the number of {\bf k}- 
and {\bf q}-points, $\Omega_{BZ}$ the volume of the BZ. 
For infinitely dense grids 
of {\bf k}- and {\bf q}-points, 
convergence of the integrand is achieved when $\sigma$ approaches zero. 
For finite grids, however, one has to find a range of $\sigma$ values 
yielding close results for the integrands at different grids. 
In metals like Nb the presence of Kohn anomalies in the phonon
spectra sets additional requirements for the accuracy. Thus, it is expected
that the aforementioned integrands have to be calculated at very dense
{\bf k}-point and {\bf q}-point grids.

\begin{figure}
\epsfxsize=8cm
\centerline{
\includegraphics[scale=0.34,angle=-90]{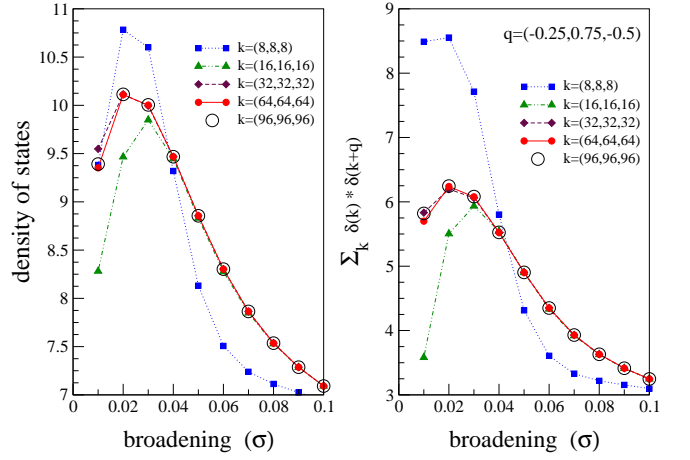}}
\caption{The density of states and the double-delta integrand on the Fermi surface of Nb.}
\label{DOS}
\end{figure}
\begin{figure}
\epsfxsize=8cm
\centerline{
\includegraphics[scale=0.34,angle=-90]{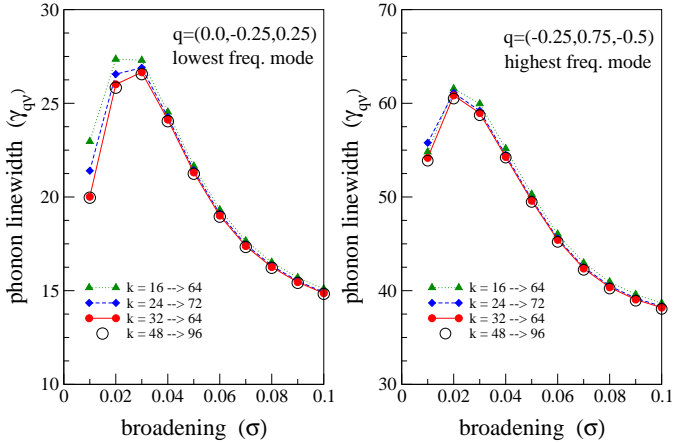}}
\caption{The phonon linewidth $\gamma_{{\bf q}\nu}$ for two selected {\bf q}-vectors.}
\label{q2q4}
\end{figure}
\begin{figure}
\epsfxsize=8cm
\centerline{
\includegraphics[scale=0.40,angle=0]{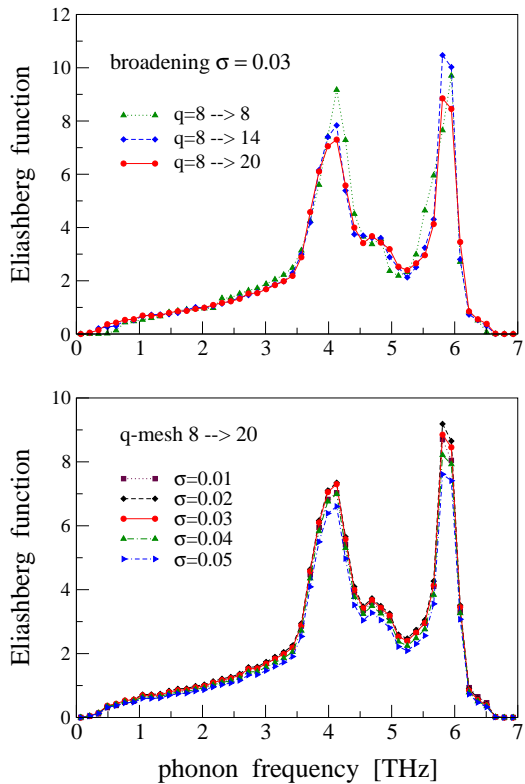}}
\vspace{-3mm}
\caption{The Eliashberg function of Nb for fixed broadening $\sigma$=0.03~Ry 
and different q-grid (top panel) and the Eliashberg function for 
fixed q-grid and different broadening $\sigma$ (bottom panel).}
\label{spectra}

\end{figure}
\begin{figure}
\epsfxsize=8cm
\centerline{
\includegraphics[scale=0.30,angle=-90]{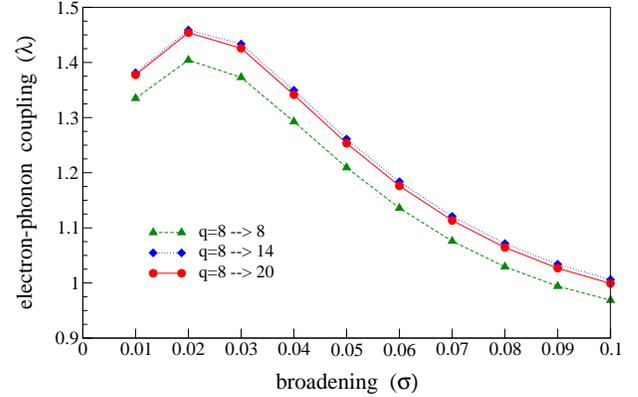}}
\caption{The total electron-phonon coupling $\lambda$ of Nb for different 
q-grids as a function of
broadening $\sigma$.}
\label{lambda-tot}
\end{figure}
\begin{figure}
\epsfxsize=8cm
\centerline{
\includegraphics[scale=0.25,angle=-90.0]{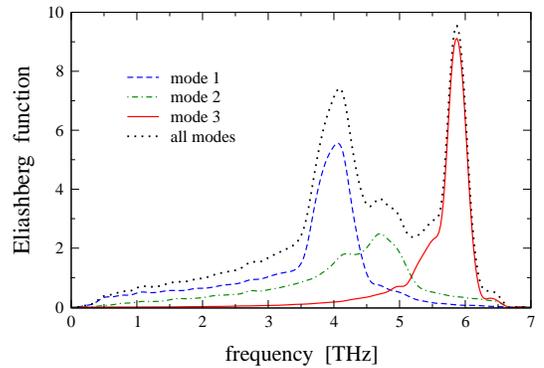}}
\caption{The Eliashberg function for Nb at calculated equilibrium lattice constant (6.14~a.u.).}
\label{a2F-0}
\end{figure}

Since the electron-phonon coupling matrix elements are smooth functions
of {\bf k} and {\bf q}, we resort to an interpolation procedure.
For a chosen {\bf k}- and {\bf q}-vector grid, we calculate matrix
elements $g_{\bf k+q, k}^{{\bf q}s\alpha,mn}$, defined as in Eq. (\ref{gg})
but with respect to the displacement of a single atom $s$ along
cartesian component $\alpha$. For each {\bf q}-vector we make a 
linear interpolation in {\bf k}-space of the matrix elements to a 
denser {\bf k}-vector grid. We then perform the integration, using 
the denser grid and gaussian broadening as in Eq. (\ref{integral}), 
of auxiliary phonon linewidths 
$\widetilde\gamma^{\alpha\beta}_{ss'}({\bf q})$, defined as:
\begin{eqnarray}
\widetilde\gamma^{\alpha\beta}_{ss'}({\bf q}) & = & \sum_{mn}\sum_{{\bf k}} 
(g_{{\bf k+q, k}}^{{\bf q}s\alpha, mn})^*
g_{{\bf k+q, k}}^{{\bf q}s'\beta, mn} \; \nonumber \\
&& \times \delta (\varepsilon_{{\bf k}+{\bf q},m}-\varepsilon_{F}) 
\delta (\varepsilon_{{\bf k},n}-\varepsilon_{F}).
\label{auxgamma}
\end{eqnarray}
These are related to the $\gamma_{{\bf q}\nu}$ of Eq. (\ref{eq5}) 
through the relation
\begin{equation}
\gamma_{{\bf q}\nu} = 2\pi\omega_{{\bf q}\nu} \sum_{s\alpha}\sum_{s'\beta} 
(u^{{\bf q}\nu}_{s\alpha})^*\widetilde \gamma^{\alpha\beta}_{ss'}({\bf q}) 
u^{{\bf q}\nu}_{s'\beta}.
\end{equation}
Symmetry is exploited to reduce the number of {\bf k}- and
{\bf q}-points used in the calculation. A convenient way to achieve 
such a goal is to perform {\em symmetrization}. Let us denote
with $T^{\alpha \beta}$ the symmetry operators of the small group 
of {\bf q} (i.e. the subgroup of crystal symmetry that leaves {\bf q}
unchanged). We restrict summation on {\bf k}-points to the irreducible
BZ calculated with respect to the small group of {\bf q}. Then
symmetrization is performed, separately for each {\bf q}-vector,
as follows:
\begin{eqnarray}
\widetilde \gamma^{\alpha \beta}_{ss'}({\bf q}) & = & 
\sum_{\alpha' \beta'} \; T^{\alpha \alpha'} T^{\beta \beta'}
\widetilde\gamma^{\alpha' \beta'}_{T(s)T(s')}({\bf q})  \nonumber \\
 & & \times exp \left( i \vec{\bf q}\cdot
     ( \vec{\tau}_{s}-\vec{\tau}_{T(s)}-\vec{\tau}_{s'}+\vec{\tau}_{T(s')}) \right).
\label{symmetrization}
\end{eqnarray}
In the above formula, atom $s$ with atomic position $\vec{\tau}_s$
transform into atom $T(s)$  with atomic position $\vec{\tau}_{T(s)}$ 
after application of operation $T$. 
The symmetrized matrix at each {\bf q} is subsequently rotated, 
using the remaining crystal symmetries that are not in the small 
group of {\bf q}, and the symmetrized matrices at all {\bf q}-vectors
in the star of {\bf q} are thus obtained with minimal computational
effort. The same procedure can be applied to dynamical matrices,
Eq. (\ref{eq11}).

Once the electron-phonon coupling matrix of Eq. (\ref{auxgamma})
are calculated on a {\bf q}-vector grid, it is possible to perform
Fourier interpolation and to interpolate to a finer grid. In this way,
the integration in the {\bf q} space needed to calculate 
$\alpha^{2}F(\omega)$, Eq. (\ref{a2F}), and $\lambda$, Eq. (\ref{lambda}),
can be accurately performed with a reasonable computational effort.

Let us turn now to some numerical experiments.
FIG.~\ref{DOS} shows the density of states for Nb (left panel)
and the double-delta integrand of a constant function (right panel)
at the Fermi surface for a selected {\bf q}-vector, as a function 
of the broadening $\sigma$ and of the Monkhorst-Pack\cite{mesh}
{\bf k}-point grid.  For large enough $\sigma$ the results 
for different grids -- except the (8,8,8) grid which is too coarse -- 
converge to the same values, which however depend on $\sigma$.
For small enough $\sigma$ different grids yield different results. 
Since we are interested in the $\sigma\rightarrow 0$ limit, we have 
to choose a grid of an affordable size that yields converged results 
for a $\sigma$ as small as possible. A reasonable choice is the 
(64,64,64) grid with $\sigma$=0.02~Ry, yielding a DOS at the Fermi 
energy about 10.1 states per spin and per Ry.
The convergence of the double-delta function is a little bit slower 
than that of a single-delta. 

The phonon linewidths, $\gamma_{{\bf q}\nu}$, for two selected phonons
{\bf q}$\nu$ are displayed in FIG.~\ref{q2q4}.  The self-consistent
calculations were performed i) at the {\bf k}-grids of (16,16,16) and
(32,32,32), interpolated to a denser (64,64,64) grid; ii)
at the (24,24,24) {\bf k}-grid, interpolated to (72,72,72); iii)
at the (48,48,48) {\bf k}-grid, interpolated to (96,96,96). 
The integration weights for the {\bf k}-space quadrature,
i.e. the gaussians centered around the single-particle energies, were
obtained from the accurate self-consistent calculation at the
corresponding dense grids.  As one can see in FIG.~\ref{q2q4}, the
convergence in {\bf k}-points is obtained quite easily even for the
SCF calculation at the grid (16,16,16).

In order to obtain the spectral function $\alpha^{2}F(\omega)$ one
needs to perform the {\bf q}-space quadrature of the phonon
linewidths; Eq. (\ref{a2Feq}). For the phonon and the electron-phonon
densities of states at given frequency $\omega$, we employ the
tetrahedron method within the scheme proposed by
Bl\"ochl.\cite{Bloechl} FIG.~\ref{spectra} (upper panel) shows the
Eliashberg function, $\alpha^{2}F$, of Nb for {\bf q}-grid (8,8,8),
without interpolation and interpolated into denser grids.
This has been done for a fixed broadening, $\sigma$=0.03~Ry.  The
lower panel of the same figure shows the $\alpha^{2}F$ function at the
{\bf q}-grid of (8,8,8) interpolated to (20,20,20)-point grid for
several broadenings $\sigma$.

In FIG.~\ref{lambda-tot}, we report the variation of total
electron-phonon coupling constant $\lambda$ with the broadening
$\sigma$ for three {\bf q}-meshes.

In FIG.~\ref{a2F-0}, we present the electron-phonon density of states
for niobium under ambient pressure.  We fit the curve of the spectral
function with cubic splines \cite{NR} for a finer plot.

%% file: main.bbl
\begin{thebibliography}{100}
\bibitem{Struzhkin} V.~V.~Struzhkin, Y.~A.~Timofeev, R.~J.~Hemley and {H.-k.}~Mao, 
{Phys.~Rev.~Lett.} {\bf 79}, 4262 (1997).
\bibitem{Ostanin} S.~A.~Ostanin, V.~Yu.~Trubitsin, S.~Yu.~Savrasov, M.~Alouani and H.~Dreyss\'e,
{Comp.~Mat.~Sc.} {\bf 17}, 202 (2000).
\bibitem{Ma} J.~S.~Tse, Z.~Li, K.~Uehara, Y.~Ma and R.~Ahuja, {Phys.~Rev.~B} {\bf 69}, 132101 (2004).
\bibitem{Anderson-1} J.~R.~Anderson, D.~A.~Papaconstantopoulos, J.~W.~McCaffrey and J.~E.~Shirber,
{Phys.~Rev.~B} {\bf 7}, 5115 (1973).
\bibitem{Anderson-2} J.~R.~Anderson, D.~A.~Papaconstantopoulos and J.~E.~Shirber,
{Phys.~Rev.~B} {\bf 24}, 6790 (1981).
\bibitem{anomalies} W.~Kohn, {Phys.~Rev.~Lett.} {\bf 2}, 393 (1959).
\bibitem{anomalies-exp} A.~D.~B.~Woods and S.~H.~Chen, {Solid~State~Commun.} {\bf 2}, 
233 (1964);A.~D.~B.~Woods, {Phys.~Rev.} {\bf 136}, A781 (1964); 
Y.~Nakagawa and A.~D.~B.~Woods, {Phys.~Rev.~Lett.}
{\bf 11}, 271 (1963); J.~Peretti, I.~Pelah and W.~Kley, {Phys.~Lett.} {\bf 3}, 105 (1962).
\bibitem{DFPT} S.~Baroni, P.~Giannozzi and A.~Testa, {Phys.~Rev.~Lett.} {\bf 58}, 1861 (1987).
\bibitem{Review} S.~Baroni, S.~de~Gironcoli, A.~Dal~Corso and {P.~Giannozzi},
{Rev.~Mod.~Phys.} {\bf 73}, 515 (2001).
\bibitem{Vanderbilt} D.~Vanderbilt, {Phys.~Rev.~B} {\bf 41}, R7892 (1990).
\bibitem{AllenGamma} P.~B.~Allen, {Phys.~Rev.~B} {\bf 6}, 2577 (1972).
\bibitem{DFT} P.~Hohenberg and W.~Kohn, {Phys.~Rev.} {\bf 136}, {B864} (1964);
W.~Kohn and L.~J.~Sham, {Phys.~Rev.} {\bf 140}, {A1133} (1965).
\bibitem{HBS1} D.~R.~Hamann, M.~Schl\"uter and C.~Chiang, {Phys.~Rev.~Lett.} {\bf 43}, 1494 (1979);
G.~B.~Bachelet, D.~R.~Hamann and M.~Schl\"uter, {Phys.~Rev.~B} {\bf 26}, 4199 (1982).
\bibitem{KB} L.~Kleinman and D.~M.~Bylander, {Phys.~Rev.~Lett.} {\bf 48}, 1425 (1982);
D. L. Bylander and L. Kleinman, {Phys.~Rev.~B} {\bf 43}, 12070 (1991).
\bibitem{MT} N.~Troullier and J.~L.~Martins, {Phys.~Rev.~B} {\bf 43}, 1993 (1991).
\bibitem{AndreaUS} A.~Dal~Corso, {Phys.~Rev.~B} {\bf 64}, 235118 (2001); A.~Dal~Corso, A.~Pasquarello,
A.~Baldereschi, {Phys.~Rev.~B} {\bf 56}, 11372 (1997).
\bibitem{GGA} A.~Dal Corso and S.~de~Gironcoli, {Phys.~Rev.~B} {\bf 62}, 273 (2000).
\bibitem{Paxton} M.~Methfessel and A.~T.~Paxton, {Phys.~Rev.~B} {\bf 40}, 3616 (1989).
\bibitem{SdG-broad} S.~de~Gironcoli, {Phys.~Rev.~B} {\bf 51}, 6773 (1995).
\bibitem{mesh} H.~J.~Monkhorst and J.~D.~Pack, {Phys.~Rev.~B} {\bf 13}, 5188 (1976).
\bibitem{espresso} http://www.quantum-espresso.org
\bibitem{Ashcroft}
R.~L.~Barns, {J.~Appl.~Phys.} {\bf 39}, 4044 (1968);
Landolt-B\"ornstein, {\em Numerical Data and Functional Relationships in Science and Technology},
edited by O. Madelung, Group III: {\em Crystal and Solid State Physics}, vol. 14: {\em Structure Data
of Elements and Intermetallic Phases}, (Springer Verlag, Berlin 1988).
\bibitem{bulkmod} A~M.~James and M.~P.~Lord,
{\em Macmillan's Chemical and Physical Data}, (Macmillan, London, UK, 1992).
\bibitem{xcrysden} A.~Kokalj, Comp. Mater. Sci., 2003, Vol. 28, p. 155. 
Code available from http://www.xcrysden.org/.
\bibitem{Solanki} A.~K.~Solanki, R.~Ahuja and S.~Auluck, {Phys.~Stat.~Sol.~B} {\bf 162}, 497 (1990).
\bibitem{lambda-1.52} L.~L.~Boyer, D.~A.~Papaconstantopoulos and B.~M.~Klein {Phys.~Rev.~B} {\bf 15},
3685 (1997).
\bibitem{Sav-lambda} S.~Y.~Savrasov and D.~Y.~Savrasov, {Phys.~Rev.~B} {\bf 54}, 16487 (1996).
\bibitem{Bauer} R.~Bauer, A.~Schmid, P.~Pavone and D.~Strauch, {Phys.~Rev~B} {\bf 57}, 11276 (1998).
\bibitem{Singalas} M.~M.~Singalas and D.~A.~Papaconstantopoulus, {Phys.~Rev.~B} {\bf 50}, 7255 (1994).
\bibitem{Crabtree} G.~W.~Crabtree, D.~H.~Dye, D.~P.~Karim and D.~D.~Koelling, {Phys.~Rev.~Lett.}
{\bf 42}, 390 (1979).
\bibitem{lambda-exp1} E.~L.~Wolf, {\em Principles of Electronic Tunneling Spectroscopy}, (Oxford
University Press, New York, 1985).
\bibitem{lambda-exp2} M.~J.~Bostock, M.~L.A.~Mac Vicar, G.~B.~Arnold, J.~Zasadzi\'nski and
E.~L.~Wolf, in {\em Proceedings of the Third International Conference on Superconductivity of d- and
f-Band Metals}, edited by H.~Suhl and M.~B.~Maple (Academic Press, New York, 1980), p. 153;
M.~J.~Bostock, V.~Diadiuk, W.~N.~Cheung, K.~H.~Lo, R.~M.~Rose and M.~L.A.~Mac Vicar, {Phys.~Rev.~Lett.}
{\bf 36}, 603 (1976).
\bibitem{Vonsovsky} S.~V.~Vonsovsky, Yu.~A.~Izyumov and E.~Z.~Kumaev, {\it Superconductivity of
Transition metals}, Springer Series in Solid State Sciences Vol. 27 (Springer-Verlag, Berlin, 1982),
Sec. 2.3.5, p. 55 and Sec. 3.9.2, p. 174.
\bibitem{fermi-1} N.~Elyashar and D.~D.~Koelling, {Phys.~Rev.~B} {\bf 15}, 3620 (1977);
{\bf 13}, 5362 (1976).
\bibitem{fermi-2} S.~Wakoh, Y.~Kubo and J.~Yamashita, {J.~Phys.~Soc.~Jpn.} {\bf 38} 416 (1975).
\bibitem{Neve} J.~Neve, B.~Sundqvist and \"O.~Rapp, {Phys.~Rev.~B} {\bf 28}, 629 (1983).
\bibitem{fermi-3} D.~P.~Karim, J.~B.~Ketterson and G.~W.~Crabtree, {J.~Low~Temp.~Phys.} {\bf 30}, 389 (1978).
\bibitem{fermi-4} S.~Shiotani, T.~Okada and T.~Mizoguchi, {J.~Phys.~Soc.~Jpn.} {\bf 38} 423 (1975).
\bibitem{broadening} K.-M.~Ho, C.-L.~Fu, B.~N.~Harmon, W.~Weber and D.~R.~Hamann, {Phys.~Rev.~Lett.}
{\bf 49}, 673 (1982); C.-L.~Fu and K.-M.~Ho, {Phys.~Rev.~B} {\bf 28}, 5480 (1983);
R.~J.~Needs, R.~M.~Martin and O.~H.~Nielsen, {Phys.~Rev.~B} {\bf 33}, 3778 (1986).
\bibitem{tetra} O.~Jepsen and O.~K.~Andersen, {Solid state. Commun.} {\bf 9}, 1763 (1971);
G.~Lehmann and M.~Taut, {Phys. Status Solidi B} {\bf 54}, 469 (1972).
\bibitem{Bloechl} P.~Bl\"ochl, O.~Jepsen and O.~K.~Andersen, {Phys.~Rev.~B}
{\bf 49}, 16223 (1994).
\bibitem{NR} {\em Numerical Recipies in Fortran}, edited by W. H. Press,  
S.~A.~Teukolsky, W.~T.~Vetterling, and B.~P.~Flannery - sec. ed.
(Cambridge University Press 1986 1992).
\end{thebibliography}
